\documentclass[runningheads]{llncs}
\usepackage[T1]{fontenc}
\usepackage{xcolor}
\usepackage{graphicx}
\usepackage{boogie}
\usepackage{eiffel}
\usepackage{listings}
\usepackage{graphicx}
\usepackage{url}
\usepackage{makecell}
\usepackage{paralist}
\usepackage{subfigure}
\usepackage{comment}

\usepackage{bbding}
\usepackage[]{graphicx}
\usepackage{scalerel}
\usepackage{algpseudocode}
\usepackage{algorithm}
\usepackage{amsmath}
\usepackage{colortbl}
\usepackage{hyperref}
\usepackage{soul}
\usepackage{booktabs}
\usepackage{array}
\usepackage{adjustbox}
\usepackage{colortbl}
\usepackage[flushleft]{threeparttable}
\usepackage{caption}
\usepackage{color}

\sethlcolor{lightgray}
\newcommand{\e}[1]{\mbox{\lstinline[basicstyle=\normalsize]|#1|}}

\usepackage{amsfonts, amsmath, amssymb}

\usepackage{fontawesome}

\begin{document}
\title{Seeding Contradiction: a fast method for generating full-coverage test suites}
\author{
Li Huang\inst{1} \and
Bertrand Meyer\inst{1, 2}\orcidID{0000-0002-5985-7434} \and
Manuel Oriol \inst{1}
}
\authorrunning{L. Huang, B. Meyer and M. Oriol}
\titlerunning{Seeding Contradiction: fast generation of full-coverage test suites}%
\institute{Constructor Institute, Schaffhausen, Switzerland\\
\email {\{li.huang, bm, mo\}@constructor.org}
\url{https://constructor.org}
\and
Eiffel Software, Santa Barbara, California\\
\url{https://eiffel.com}}
\maketitle              % typeset the header of the contribution
\begin{abstract}

The regression test suite, a key resource for managing program evolution, needs  to achieve 100\% coverage, or very close, to be useful.
Devising a test suite manually is unacceptably tedious, but existing automated methods are often inefficient. The method described in this article, ``Seeding Contradiction'', inserts incorrect instructions into every basic block of the program, enabling an SMT-based Hoare-style prover to generate a counterexample for every branch of the program and, from the collection of all such counterexamples, a test suite. The method is static, works fast, and achieves excellent coverage.

\end{abstract}
\textit{Draft of article to be presented at ICTSS 2023 (International Conference on Testing Software and Systems) in Bergamo, 18-20 September 2023.} 

\section{Overview} \label{Introduction}
In the modern theory and practice of software engineering, tests have gained a place of choice among the artifacts of software production, on an equal footing with code. One particularly important rule is that every deployed program should come accompanied with a \textit{regression test suite} achieving high branch coverage and making it possible to check, after any change to the software, that previous functionality still works: no ``regression'' has occurred.

Producing a high-coverage regression test suite is a delicate and labor-intensive task. Tools exist (RANDOOP \cite{pacheco2007randoop}, Pex \cite{tillmann2008pex}, AutoTest \cite{autotest}, Korat \cite{boyapati2002korat}) but they are typically \textit{dynamic}, meaning that they require numerous executions of the code. The Seeding Contradiction (SC) method and supporting tools presented in this article typically achieve 100\% coverage (excluding unreachable code, which they may help detect) and involve no execution of the code, ensuring very fast results.

The principal insight of Seeding Contradiction is to exploit the power of modern program provers, which attempt to generate a counterexample of program correctness. In normal program proving, we hope that the prover will \textit{not} find such a counterexample: a proof follows from the demonstrated \textit{inability} to \textit{disprove} the program’s correctness. Switching the focus from proofs to tests, we may look at counterexamples in a different way: as test cases. We may call this approach \textit{Failed Proofs to Failing Tests} or FP-FT. Previous research (including by some of the present authors) has exploited FP-FT in various ways \cite{nilizadeh2022generating} \cite{huang2022failed} \cite{huang2022improving}. Seeding Contradiction  extends FP-FT to a new goal: generating a full-coverage test suite, by applying FP-FT to \textit{seeded} versions of the program in which a branch has on purpose been made incorrect. For every such variant, the prover generates a counterexample exercising the corresponding branch. Combining the result for all branches yields a high-coverage test suite. In fact coverage is normally 100\%, with the following provisions:
\begin{itemize}[$\bullet$]
    \item Some branches may be unreachable. Then by definition no test could cover them; the tool may help identify such cases. (Terminology: we will use the term \textbf{exhaustive coverage} to mean 100\% coverage of reachable branches.)
    \item Limitations of the prover may prevent reaching 100\%. In our examples so far such cases do not arise.
\end{itemize}

\noindent The method involves no execution of the code and on examples tried so far produces a test suite much faster than dynamic techniques (section \ref{evaluation}).

The current setup involves the AutoProof \cite{tschannen2015autoproof} \cite{autoproof} verification framework for contract-equipped Eiffel \cite{meyer1997object} code, relying internally on the Boogie proof system \cite{leinoboogie} \cite{barnett2005boogie} and the Z3 SMT solver \cite{de2008z3}. It is generalizable to other approaches.

The discussion is organized as follows. Section \ref{method} presents the approach by considering a small example. Section \ref{correctness} examines the theoretical correctness of that approach. Section \ref{implementation} describes the extent to which we have applied it so far, and section \ref{evaluation} assesses the results. Section \ref{limitations} discusses limitations of the current state of the work and  threats to validity of the evaluation results. Section \ref{related} reviews related work and section \ref{conclusion} presents conclusions and future work.

\section{The method} \label{method}
A simple code example will illustrate the essential idea behind Seeded Composition.

\subsection {Falsifying a code block} \label {falsify}
Consider a small routine consisting of a single conditional instruction:
\begin{lstlisting}[language = Eiffel, basicstyle=\fontsize{0.33cm}{0.33cm}]
    simple (a: INTEGER)
        do
            if a > 0 $ $ then  x := 1 $ $ else  x := 2 $ $ end
        end
\end{lstlisting}
where \e{x} is an integer attribute of the enclosing class. In a Design-by-Contract approach intended to achieve correctness by construction, the routine might include the following postcondition part (with $\Longrightarrow$ denoting implication):
\begin{lstlisting}[language = Eiffel, basicstyle=\fontsize{0.33cm}{0.33cm}]
    ensure
        a > 0 $ $ $\Longrightarrow$ x = 1
        a <= 0 $ $ $\Longrightarrow$ x = 2
\end{lstlisting}
\noindent With or without the postcondition, how can we obtain a regression test suite that will exercise both branches?

\begin{figure}[htbp]
\vspace{-0.4cm}
\centerline{{\includegraphics[width=2.4in]{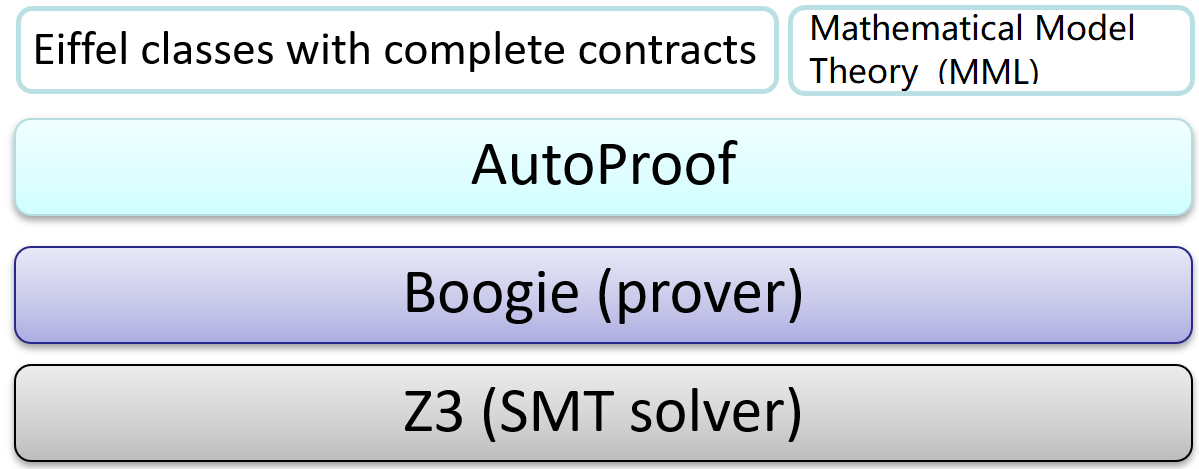}
}}
\vspace{-0.2cm}
\caption{AutoProof tool stack}
\label{fig: tool stack}
\vspace{-0.6cm}
\end{figure}

Various techniques exist, discussed in section \ref{related} and generally requiring execution of the code. The Seeding Contradiction technique is, as noted, static (it does not involve executing the code); it assumes that we have a toolset for proving program correctness. Specifically, we rely on the AutoProof environment \cite{tschannen2015autoproof} \cite{autoproof}, with a tool stack presented in Fig. \ref{fig: tool stack}, in which the Boogie prover is itself based on an SMT solver, currently Z3. A characteristic of this style of proof is that it relies on a \textit{disproof} of the \textit{opposite} property: the SMT solver tries to construct at least one counterexample, violating the desired result. If it cannot find one, the proof is successful.

\noindent In this work, as in previous articles using the general FP-FT approach \cite{huang2022failed} \cite{huang2022improving}, we are interested in a proof that actually fails: then the counterexample can be useful on its own, yielding a directly usable test.

In contrast with the earlier FP-FT work, the proof that will fail is not a proof of the actual program but of a modified version, into which we have inserted (``seeded'') incorrect instructions. In the example, we change the first branch, so that the routine now reads
\vspace{-0.1cm}
\begin{lstlisting}[language = Eiffel, basicstyle=\fontsize{0.33cm}{0.33cm}]
    simple (a: INTEGER)
        do
            if a > 0 $ $ then
                check False end             -- This is the added instruction
                x := 1 $ $ -- The rest is unchanged.
            else
                x := 2
            end
        end
\end{lstlisting}
\vspace{-0.1cm}

\noindent A ``\e{check C end}'' instruction (\e{assert C} in some other notations \cite{leino2023program}) states that the programmer expects condition \e{C} to hold at the corresponding program point. Specifically, its semantics is the following, from both a dynamic perspective (what happens if it gets executed) and a static, proof-oriented perspective:
\begin{itemize}[$\bullet$]
    \item
From a dynamic viewpoint, executing the instruction means: if condition \e{C} has value \e{True} at that  point, the \e{check}  instruction has no effect other than evaluating \e{C}; if \e{C} evaluates to \e{False} and the programmer has enabled run-time assertion monitoring, as possible in EiffelStudio, execution produces a violated-assertion exception, usually implying that it terminates abnormally.

\item
In the present discussion's static approach, the goal is to prove the program correct. The semantics of the \e{check} instruction is that it is correct if and only if the condition \e{C} alway has value \e{True} at the given program point. If the prover cannot establish that property, the proof fails.

\end{itemize}

\noindent
In a general FP-FT approach, the key property is that in the static view, if the proof fails, an SMT-based prover will generate a \textbf{counterexample}. In the Seeding Contradiction approach,  \e{C} is \e{False}: the proof \textit{always} fails and we get a counterexample exercising the corresponding branch --- exactly what we need if, as part of a regression test suite, we want a test exercising the given branch.

For the \e{simple} code seeded with a \e{check False end}, such a counterexample will, by construction, lead to execution of the first branch (\e{a > 0}) of the conditional. If we have an efficient mechanism to turn counterexamples into tests, as described in earlier work \cite{huang2022failed} \cite{huang2022improving}, we can get, out of this counterexample, a test of the original program which exercises the first branch of the conditional.

Such a generated test enjoys several interesting properties:
\begin{itemize}[$\bullet$]
 \item
    It can be produced even in the absence of a formal specification (contract elements such as the postcondition above).
    \item
    Unless the enclosing code (here the routine \e{simple}) is unreachable, the test can be produced whether the program is correct or incorrect.
    \item
    If the program is correct, the test will pass and is useful as a regression test (which may fail in a later revision of the program that introduces a bug).
    \item
    Generating it does not require any execution.
    \item
    That generation process is fast in practice (section \ref{evaluation}).
\end{itemize}

\noindent
The next sections will show how to generalize the just outlined idea to produce such tests not only for one branch as here but for  \textit{all} branches of the program, as needed to obtain an exhaustive-coverage regression test suite.

\subsection {Block variables} \label{variables}

To generalize the approach, the following terminology is useful. So far it has been convenient to talk informally of ``branches'', but the more precise concept is \textbf{basic block}, defined in the testing and compilation literature as a sequence of instructions not containing conditionals or loops. (This definition is for a structured program with no branching instructions. In a more general approach, a basic block is any process node --- as opposed to decision nodes --- in the program's flowchart.) ``Block'' as used below is an abbreviation for ``basic block''.

The method illustrated on the \e{simple} example generates a test guaranteed to exercise a specific block of a correct program: seed the program by adding to the chosen block one  \e{check False end} instruction. Then, as seen in the example, we run the prover and apply the FP-FT scheme: since the program is now incorrect, the proof fails and the prover generates a counterexample, which we turn into a runnable test guaranteed to exercise the given block in the original program.

To generalize this approach so that it will generate a test suite exercising all blocks, a straightforward idea is ``\textit{Multiple Seeded Programs}'' (MSP): generate such a seeded program for each of its blocks in turn; then run the prover on every such program, in each case producing a counterexample and generating a test from it. Subject to conditions in section \ref{correctness} below, the MSP approach is correct, in the sense that together the generated tests exercise all reachable blocks. It is, however, impractical: for a single original program, we would need to generate a possibly very large number of seeded programs, and run every one of them through the prover.

To obtain a realistic process, we can instead generate a single seeded program, designed to produce the same counterexamples as would all the MSP-generated programs taken together. A helpful property of a good counterexample-based prover is that it can deal  with a program containing several faults and generate a set of counterexamples, each addressing one of the faults. In the example above, we can submit to the prover a single seeded program of the form
\vspace{-0.1cm}
\begin{lstlisting}[language = Eiffel, basicstyle=\fontsize{0.33cm}{0.33cm}]
    simple (a: INTEGER)
        do
            if a > 0 $ $ then
                check False end
                x := 1 $ $                   -- Instruction 1
            else
                check False end
                x := 2     $ $              -- Instruction 2
            end
        end
\end{lstlisting}
\vspace{-0.1cm}
\noindent which will produce two counterexamples, one for each branch. We call this approach ``RSSP'' (Repeatedly Seeded Single Program). With AutoProof, the FP-FT tools generate tests with \e{a} = 1 and \e{a} = 0. (More precisely, the prover initially generates larger and less intuitive values, but a minimization technique described in earlier work \cite{huang2022improving} produces 1 and 0.)

This approach does not suffice for more complex examples. Assume that after the conditional instruction the routine \e{simple} includes another conditional:
\begin{lstlisting}[language = Eiffel, basicstyle=\fontsize{0.33cm}{0.33cm}]
        -- This code comes after the above conditional (Instructions 1-2)
    if a$^2$  > a $ $ then
        x := 3  $ $                         -- Instruction 3
    else
        x := 4  $ $                         -- Instruction 4
    end
\end{lstlisting}

\noindent With the program seeded as above, even if we insert a \e{check False end} into each of the two new blocks (before Instructions 3 and 4), we will get tests covering only two cases (1-4, 2-4), not four (1-3, 1-4, 2-3, 2-4) as needed. These two tests, \e{a}\ = 1 and  \e{a} = 0, fail to cover Instruction 3. The reason is that the prover does not generate specific tests for the branches of the second conditional (3-4) since it correctly determines that they are unreachable as both branches of the first conditional (1-2) now include a \e{check False end}. They were, however, both reachable in the original! The test suite fails to achieve exhaustive coverage.

%\noindent \li{With the program seeded as above, even if we insert a \e{check False end} into each of the two new blocks, we will only get two tests cases: \e{a} = 1 that cover instructions 1 and 4, and \e{a} = 0 covering instructions 2 and 4. There is no test covering the block of instruction 3. It is not guaranteed that the counterexamples will cover both branches of the second conditional (Instructions 3 and 4), since the prover correctly determines that they are unreachable, as both branches of the first conditional have been seeded with a \e{check False end}.}

The solution to this ``\textit{Seeded Unreachability}'' issue is to make the \e{check} themselves conditional. In the seeded program, for every routine under processing, such as \e{simple}, we may number every basic block, from 1 to some \e{N}, and add to the routine  an argument $bn$ (for ``block number'') with an associated precondition
\vspace{-0.5cm}
\begin{lstlisting}[language = Eiffel, basicstyle=\fontsize{0.33cm}{0.33cm}]
    require
        $bn$ $\geq$ 0               $         $       -- See below why 0 $ $ and not 1.
        $bn$ $\leq$  N

\end{lstlisting}
\vspace{-.05cm}
\noindent To avoid changing the routine's interface (as the addition of an argument implies), we will instead make $bn$ a local variable and add an initial instruction that assigns to $bn$, non-deterministically, a value between 0 and \e{N}. Either way, we now use, as seeded instructions, no longer just \e{check False end} but
\begin{lstlisting}[language = Eiffel, basicstyle=\fontsize{0.33cm}{0.33cm}]
    if $bn = i$ then check False end end
\end{lstlisting}
\vspace{-.1cm}
\noindent where $i$ is the number assigned to the block. In the example, the fully seeded routine body for the extended version of \e{simple}  with two conditionals, is (choosing the option of making \e{bn} a local variable rather than an argument):
\vspace{-.2cm}
\begin{lstlisting}[language = Eiffel, basicstyle=\fontsize{0.33cm}{0.33cm}]

   bn := ``Value chosen non-deterministically between 0 $ $ and N''
   if a > 0 $ $ then
        if $bn = 1$ $ $ then  check False end end
        x := 1 $ $                   -- Instruction 1
    else
        if $bn = 2$ $ $ then  check False end end
        x := 2     $ $              -- Instruction 2
    end

    if a$^2$  > a $ $ then
        if $bn = 3$ $ $ then check False end end
        x := 3  $ $                           -- Instruction 3
    else
        if $bn = 4$ $ $ then check False end end
        x := 4  $ $                          -- Instruction 4
    end

\end{lstlisting}
\vspace{-.1cm}
\noindent
As in the previous attempt, there are four incorrect \e{check False} instructions, but all  are now reachable for $bn$ values ranging from 1 to 4. The prover generates counterexamples exercising all the paths of the original program (with appropriately generated values for its original variables). In this case there is only one relevant variable, $a$; AutoProof's prover generates, for the pair $[bn, a]$,
%\li{actually, the test input only contain the value of a?}
the test values [1, 1], [2, 0], [3, -1], [4, 0]. These four tests provide 100\% branch coverage for the program and can serve as a regression test suite. We call this technique \textbf{Conditional Seeding}; it addresses the Seeded Unreachability issue.

As noted above, we accept for $bn$ not only values between 1 and $N$ (the number of basic blocks) but also 0. This convention has no bearing on test generation and coverage but ensures that the behavior of the original program remains possible in the seeded version: for $bn$ = 0, none of the seeded \e{check False} will execute, so the program behaves exactly as the original. If the original was correct, the prover will not generate any counterexample for that value.

\section {Correctness} \label {correctness}
The goal of a test-suite-generation strategy is to produce high-coverage test suites. The Seeding Contradiction strategy is more ambitious: we consider it correct if it achieves \textbf{exhaustive coverage} (as defined in section \ref{Introduction}: full coverage of reachable branches). More precisely, we will now prove that SC is ``coverage-complete'' if the prover is ``reachability-sound'', ``correctness-sound'' and ``counterexample-complete''.  \ref{assumptions}  defines these concepts and \ref{proof} has the proof.
\vspace{-0.3cm}
\subsection {Definitions and assumptions} \label{assumptions}
\vspace{-0.2cm}
Establishing the correctness of SC requires precise conventions and terminology.

A general assumption is the availability of an ``FP-FT'' mechanism which, as described in previous articles~\cite{huang2022failed}, can produce directly executable tests (expressed in the target programming language, in our case Eiffel) from counterexamples produced by the SMT-based prover. As a consequence, the rest of this discussion does not distinguish between the notions of counterexample and test.\footnote{Counterexamples that the prover generates at first can use arbitrary values, sometimes too large to be meaningful to programmers; as noted in \ref{variables}, a minimization strategy is available to produce more intuitive values. The SC technique and its analysis are independent of such choices of counterexamples.}

The definition of basic block, or just \textbf{block} for short, appeared earlier (\ref{variables}).

For simplicity, we assume that the programs are \textbf{structured}, meaning that they use  sequences, loops and conditionals as their only control structures. Also, we consider that a conditional always includes exactly one ``else'' part (possibly empty), and that a loop has two blocks, the loop body and an empty block (corresponding to the case of zero iterations). Further, expressions, particularly conditional expressions used in conditional instructions, are side-effect-free. Thanks to these conventions, instruction coverage (also known as statement coverage) and branch coverage are the same concept, called just ``coverage'' from now on.

A (possibly empty) block of a program is \textbf{reachable} if at least one set of input values will cause it to be executed, and otherwise (if, regardless of the input, it cannot be executed) \textit{unreachable}.
Reachability is an undecidable property for any realistic programming language, but that need not bother us since this work relies on a prover of which we will only require that it be \textbf{reachability-sound}: if a block is reachable, the prover will indeed characterize it as reachable.
(The prover might, the other way, wrongly characterize a block as reachable when in fact it is not: with \e{if} \e{cos$^2$} \e{(x)} \e{+} \e{sin$^2$} \e{(x)} \e{=} \e{100} \e{then} \e{y}\ \e{:=}\ \e{0} \e{else} \e{y} \e{:=} \e{1} \e{end}, the prover might consider \e{y = 0} as a possible outcome if it does not have enough built-in knowledge about trigonometric functions. That too-conservative determination does not endanger the SC strategy.)

A program may contain instructions of the form \e{check C end}, with no effect on execution (as previewed in section \ref{method}). Such an instruction is \textbf{correct} if and only if the condition \e{C} will hold on every execution of the instruction. This property is again undecidable, and again we only need the prover to be \textbf{correctness-sound}: if it tells us that an instruction is correct, it is. (We hope the other way around too, but do not require it.) For the SC strategy we are interested in the trivial case for which \e{C} is \e{False}.

Also for simplicity, we assume that all correctness properties are expressed in the form of \e{check}  instructions; in particular, we replace any contract elements (preconditions, postconditions, loop invariants and variants, class invariants) by such instructions added at the appropriate places in the program text.

With this convention, a \textbf{block} is correct if all its \e{check} instructions are, and a \textbf{program} is correct if all its blocks are. For a normally written program, this definition means that the program is correct in the usual sense; in particular, if it has any contracts, it satisfies them, for example by having every routine ensure its postcondition. The SC strategy, by adding \e{check False end} to individual blocks, makes these blocks --- and hence the program as a whole --- incorrect.

A \textbf{test suite} is a collection of test cases for a program.

A test suite achieves \textbf{exhaustive coverage} if for every reachable block in the program at least one of  its test cases causes that block to be executed. (Note the importance of having a reachability-sound prover: if it could wrongly mark some reachable blocks as unreachable, it could wrongly report exhaustive coverage, which is not acceptable. On the other hand, if it is reachability-sound, it may pessimistically report less-than-exhaustive coverage for a test suite whose coverage is in fact exhaustive, a disappointing but not lethal result. This case does not occur in our examples thanks to the high quality of the prover.)

A test-suite-generation method (such as Seeding Contradiction) is \textbf{coverage-complete} if the generated test suite achieves exhaustive coverage for any correct program. In other words, for each reachable basic block of a correct program, at least one test in the suite will execute the block.

Finally, consider a prover that can generate counterexamples for programs it cannot prove correct. The prover is \textbf{counterexample-complete} if it generates a counterexample for every block that it determines to be reachable and incorrect.

With these conventions, the correctness of the Seeding Contradiction method is the property (proven next) that
\begin{itemize}[$\bullet$]
    \item []
   \textit{ If the prover is reachability-sound, correctness-sound and counterexample-complete, SC is coverage-complete.}
\end{itemize}

\subsection {Proof of correctness} \label {proof}
\noindent To establish that correctness holds, on the basis of the preceding definitions, we first establish the following two lemmas:
\begin{itemize}[$\bullet$]
    \item [1]
    Any test case of a seeded program (the program modified by addition of \e{check} instructions as described above) yields, by omitting the \e{bn} variable, a test case of the original program, exercising the same basic block.

    \item [2]
    Any reachable block of the original program is reachable in the seeded one.
\end{itemize}

\noindent The proof of both lemmas follows from the observation that the seeded program has the same variables as the original except for the addition of the \e{bn} variable, which only appears in the conditional \e{check} instructions and hence does not affect the behavior of the program other than by possibly causing execution of one of these instructions in the corresponding block. If \e{bn} has value \e{i} in such an execution, the execution of all blocks other than the block numbered \e{i} (if any --- remember that we accept the value 0 for \e{bn}), in particular the execution of any block in an execution path \textit{preceding} the possible execution of block \e{i}, proceeds exactly as in the original unseeded program. As a result:
\begin{itemize}[$\bullet$]
    \item
    Any test executing block number \e{i} in the seeded program for any \e{i} has, for all other variables (those of the original program), values that cause execution of block \e{i} in the original program too, yielding Lemma 1.
    \item
    Consider a reachable block, numbered \e{i}, of the original program. Since it is reachable, there exists a variable assignment, for the variables of the original program, that causes its execution. That variable assignment complemented by \e{bn = i} causes execution of block \e{i} in the seeded program, which is therefore reachable, yielding Lemma 2.
\end{itemize}

\noindent To prove that SC satisfies the definition of correctness (given at the end of \ref{assumptions}):
\vspace{-0.1cm}
\begin{itemize}[$\bullet$]
    \item
Assume that the original program is correct; then the only incorrect instructions in the seeded program are the added conditional \e{check} instructions (the \e{if C then check False end} at the beginning of every block).

\item
Consider an arbitrary reachable basic block \e{B}, of the original program. Because of Lemma 2, it is also reachable in the seeded program.

\item
If the prover is reachability-sound, it indeed determines that block \e{B} is (in the seeded program) reachable.

\item
If the prover is also correctness-sound,it determines that \e{B}'s seeded \e{check} instruction is incorrect, and hence (by definition) that \e{B} itself is incorrect.

\item
Then if it is counter-example-complete it will generate a counterexample that executes \e{B} in the seeded program.

\item
By Lemma 1, that counterexample yields a test that executes block \e{B} in the original program.

\item
As a consequence, by the definition of correctness above, the Seeding Contradiction strategy is correct.
\end{itemize}

\subsection {Correctness in practice} \label{practice}

To determine that SC as implemented is correct, we depend on properties of the prover: the definition assumes that the prover is reachability-sound, correctness-sound and counterexample-complete.

To our knowledge, no formal specification exists for the relevant tools in our actual tool stack (Fig. \ref{fig: tool stack}), particularly Z3 and Boogie. In their actual behavior as observed pragmatically, however, the tools satisfy the required properties.

\section{Implementation} \label{implementation}

We have implemented Seeding Contradiction strategy in the form of a new option of the AutoProof program-proving framework, called ``Full-coverage Test Generation'' (FTG) \footnote{AutoProof including the FTG option is available for download at github.com/huangl223/ES-AP-Installation.}. The implementation relies on the FP-FT \cite{huang2022failed} \cite{huang2022improving} feature of AutoProof, which enables automatic generation of failed tests from failed proofs.
The objective is to add the incorrect \e{check} instructions at the appropriate program locations so that the verification of the seeded program results in proof failures, yielding an exhaustive-coverage test suite as described above.

 Like the rest of AutoProof, seeding is modular: routine by routine. It is applied at the Boogie level, so that the Eiffel program remains untouched. The Boogie equivalent of the \e{check}  instruction is written \e{assert}. Depending on the structure of the code for a routine \e{r}, five cases arise, reviewed now.

 \ 

\noindent \textbf{A - Plain Block}. \label {plain} If the body of \e{r} includes no conditional  and hence has only one path, the SC strategy  inserts a single \e{assert false} at the beginning of the body. Verification of \e{r} results in failure of the assertion; by applying FP-FT, we obtain a valid test case of \e{r} (whose test input satisfies the precondition).

\ 

\begin{figure}[htbp]
    \centering
\vspace{-0.5cm}
\centerline{{\includegraphics[width=7.5cm]{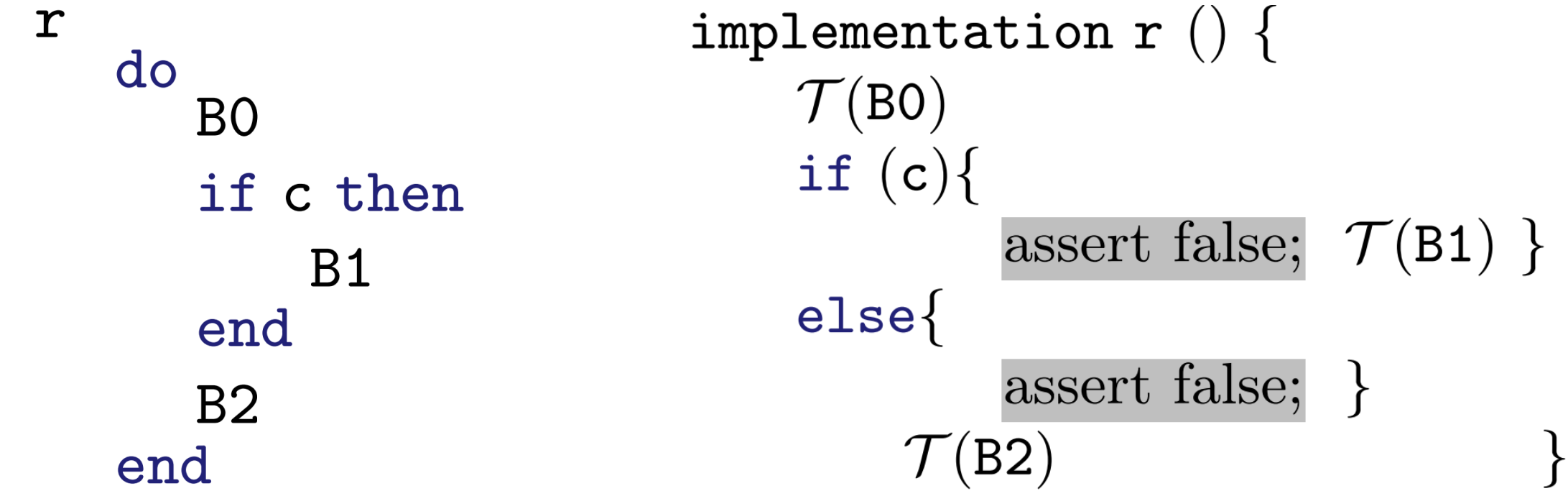}
}}
\vspace{-0.3cm}
    \caption{Instrumentation for \e{r} with implicit \e{else} branch. Left:  original Eiffel code of \e{r}. Right, seeded Boogie code.
    \e{B$_i$} ($i\in$ \{0, 1, 2\}) is a basic block in Eiffel, \e{c} a branch predicate evaluating to {\footnotesize\e{true}} or {\footnotesize\e{false}}, $\mathcal{T}$(\e{B$_i$}) the Boogie translation of \e{B$_i$}.}
    \label{listing: single_then_branch}
\end{figure}

\noindent \textbf{B - Implicit else branch}. \label{implicit} If \e{r} contains a conditional  whose \e{else} branch is implicit, SC makes it explicit and produces a test case covering the branch.
Fig. \ref{listing: single_then_branch} shows an example: SC inserts two \e{assert} clauses, one in the \e{then} branch and the other in the \e{else} branch that it creates. Running the proof produces two counterexamples for the two injected \e{assert} clauses, hence two tests.

\begin{figure}[htbp]
    \centering
\centerline{{\includegraphics[width=3.5in]{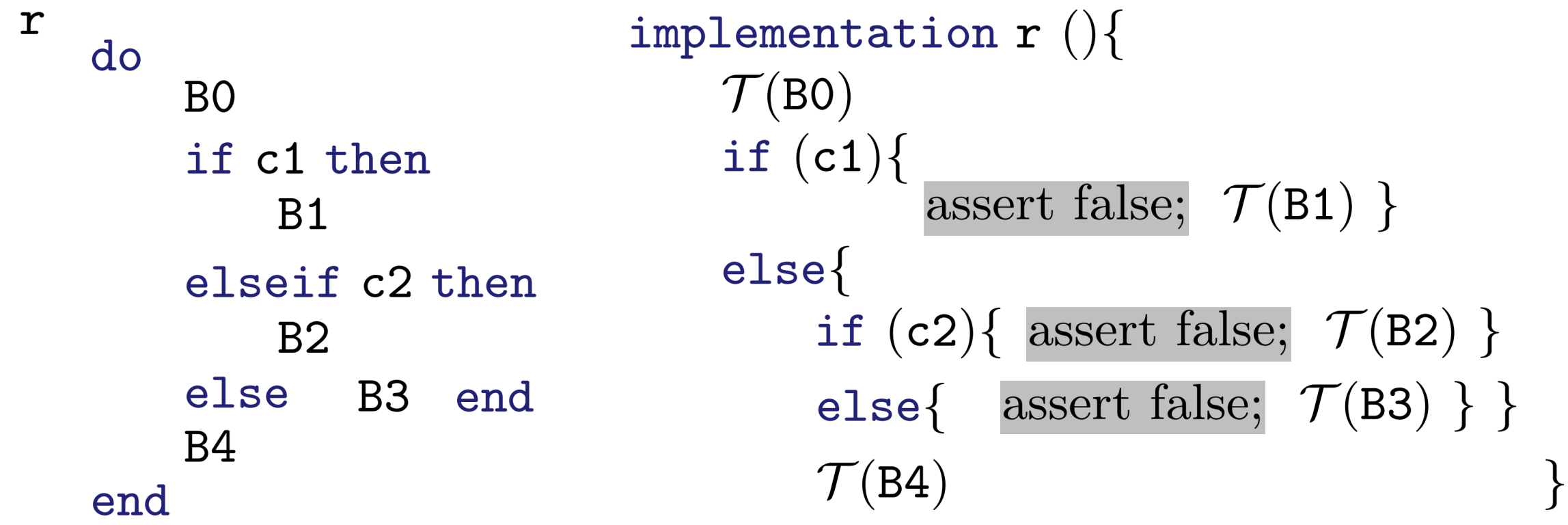}}}
    \caption{Instrumentation for cascading branches: three \e{assert false} clauses are inserted for the three branches in \e{r}; note that the \e{elseif} instruction in Eiffel, together with the last \e{else} instruction, is mapped to a nested \e{if-else} instruction in Boogie.}
    \label{listing: cascading branches}
\end{figure}

%
%%\vspace{0.1cm}
%\vspace{-0.4cm}
%\noindent \textbf{Cascading Branches}. \label{cascading} If \e{r} has a series of branches placed sequentially, to generate test cases for each branch, as in Fig. \ref{listing: cascading branches}, the SC algorithm inserts,  in the Boogie code, an \e{assert false} clause in each branch.
%Verification of the instrumented code results in failures corresponding to each of the assertions, and we can get tests whose execution would go through the assertions' locations.
%
%% why this works?
%
\noindent \textbf{C - Cascading Branches}.
\label{cascading} If \e{r} has a series of branches placed sequentially, as in Fig. \ref{listing: cascading branches}, the SC algorithm inserts an \e{assert false} clause in each branch.
The resulting tests cover all branches.

\ 

\begin{figure}[h]
    \centering
\centerline{{\includegraphics[width=8cm]{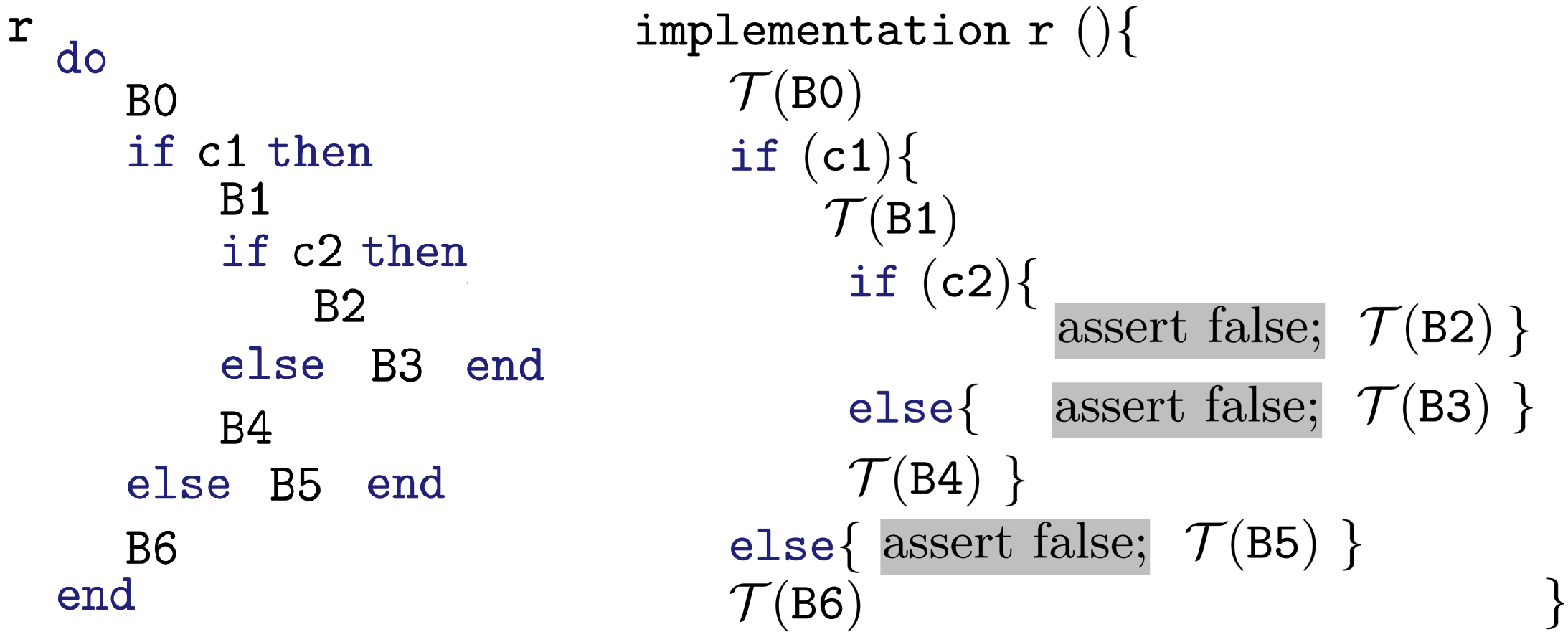}}}
    \caption{Instrumentation for nested branches}
    \label{listing: nested_branches}
\end{figure}

\noindent \textbf{D - Nested branches}. \label{nested} When conditionals are nested, SC only generates tests targeting the \textit{leaf} branches --- those with no embedded conditionals. This approach is sound since any program execution that exercises a leaf branch must also go through all the branches leading to it. Fig. \ref{listing: nested_branches} has three leaf branches for blocks \e{B2}, \e{B3} and \e{B5}. Any execution  going through \e{B2} and \e{B3} will exercise \e{B1}; SC only inserts  \e{assert}  instructions for leaves (none for \e{B1}).%reachability problem

%Unlike the cases of cascading branches, where the conditions of those branches are evaluated in succession, code execution runs whichever expression is true first, and the remaining \e{if} instructions are not checked and neither does the \e{else} branch block run.
%Inserting \e{check false} assertion would not raise reachability issue.
%

%\vspace{0.1cm}

\ 

\begin{figure}[htbp]
\vspace{-0.5cm}
    \centering
\centerline{{\includegraphics[width=9cm]{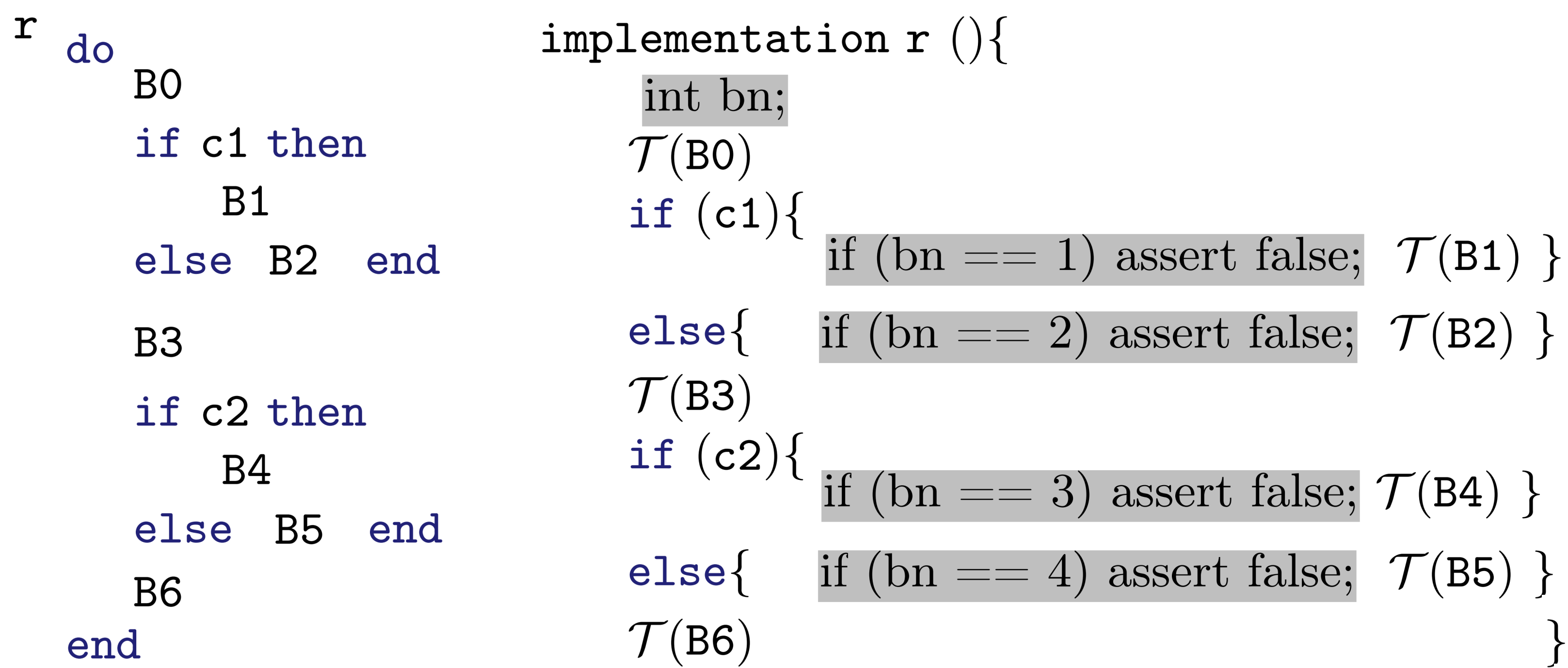}}}
\vspace{-0.3cm}
    \caption{Instrumentation for sequential conditionals}
    \label{listing: squential_decision}
\end{figure}

\noindent \textbf{E - Sequential decisions}. \label {sequential} If \e{r} has multiple successive decision instructions,
as in Fig. \ref{listing: squential_decision}, SC inserts the conditional \e{assert false} instructions as explained in \ref{variables}. It declares a variable \e{bn} for the block number and adds ``\e{if ($bn == i$)} \e{assert false;}''. Since the value of \e{bn} is between 0 and \e{N} (number of target blocks), it adds a clause ``\e{requires} \e{bn $\geq$ 0} \&\& \e{bn $\leq$ N}'' to the precondition of \e{r}.

%unreachable branches.

%1) for nested branches, only instrument the assertion into innermost branches
%2) for sequential branches, introduce $bn$ as a local variable
%3) algorithm?

%\noindent The implementation of the instrumentation is performed in the translation process.
\section{Evaluation and comparison with dynamic techniques} \label{evaluation}
We performed a performance evaluation of  Seeding Contradiction as implemented in AutoProof per the preceding section, comparing it to two existing test generation tools: IntelliTest \cite{tillmann2008pex} (previously known as Pex, a symbolic execution test-generation tool for .NET) and AutoTest \cite{autotest}, a test generation tool for Eiffel using Adaptive Random Testing, specifically ARTOO \cite{artoo}).

\vspace{-0.3cm}
\subsection{Comparison criteria and overview of the results} \label{overview}

The experiment applies all three tools to generate tests for 20 programs adapted from examples in the AutoProof tutorial\footnote{http://autoproof.sit.org/autoproof/tutorial} and benchmarks of previous software verification competitions \cite{weide2008incremental} \cite{bormer2011cost} \cite{klebanov20111st}.
Table \ref{table: examples} lists their characteristics, including implementation size (number of  Lines Of Code) and number of branches.
\begin{table}
\vspace{-1cm}
\begin{adjustbox}{width=1\textwidth}
    \begin{threeparttable}
    \caption{\textbf{Examples}}
    \begin{subtable}
     \scriptsize
  \centering
     % \vspace{-0.3cm}
   \renewcommand\arraystretch{1.3}
    \begin{tabular}{c c c c c c p{35pt} p{35pt} p{35pt} p{35pt} p{35pt} c}
    \hline
      & Account  & Clock & Heater  & Lamp & Max & Linear Search & Insertion Sort & Gnome Sort & Square root& Sum and max & Arithmetic
     \\ \hline
     \rowcolor{blue!5} LOC & 214 & 153 & 102 & 95 & 49 & 64 & 122 & 62 & 56 & 56 & 204
   \\
    Branches & 14 & 10 & 8 & 8& 3 & 5 & 5 & 5 & 5 & 4 & 14
   \\ \hline
    \end{tabular}%
  \label{table: metrics}%
\end{subtable}
   \vspace{-0.7cm}
\begin{subtable}
\centering
    \begin{tabular}{p{30pt} p{60pt} p{30pt} p{38pt} p{38pt} p{33pt} p{38pt} p{33pt} p{48pt} c}
    \hline
      Binary search & Recursive binary search & Dutch flag & Two way max & Two way sort & Quick sort & Selection Sort & Bubble Sort & Optimized gnome sort & Total
    \\ \hline
     \rowcolor{blue!5}  74 & 89 & 188 & 49 & 85 & 232 & 167 & 165 & 183 & \textbf{2409}
   \\
    5 & 7& 11 &4 & 6& 9&5&5&8 & \textbf{141}
   \\ \hline
    \end{tabular}%
   \end{subtable}
   \label{table: examples}%
    \end{threeparttable}
\end{adjustbox}
\vspace{-0.6cm}
\end{table}
%\subsection{Experiment settings}

%\begin{figure}
%    \centering
%    \includegraphics[width=3.4in]{figures/workflow.png}
%    \caption{Caption}
%    \label{fig:workflow}
%\end{figure}
\noindent The comparison addresses three metrics: coverage; time needed to generate the tests; size of the test suite. All code and results are available at \url{https://github.com/huangl223/ICTSS2023}.

The examples are originally in Eiffel; we translated them manually into C\# for  IntelliTest.
The experiment includes a test generation session for every example in every tool. For AutoTest, whose algorithms keeps generating tests until a preset time limit, it uses 10 minutes (600 seconds) as that limit; there is no time limit for the other two approaches.

All sessions took place on a machine with a 2.1 GHz Intel 12-Core processor and 32 GB of memory, running Windows 11 and Microsoft .NET 7.0.203. Versions used are: EiffelStudio 22.05 (used through AutoProof and AutoTest); Boogie 2.11.10; Z3 solver 4.8.14; Visual Studio 2022 (integrated with IntelliTest).

Table \ref{table: overview} shows an overview of the results. SC and IntelliTest handle the examples well, with coverage close to 100\%; SC reaches exhaustive coverage (100\% coverage of reachable branches) for all 20 examples and IntelliTest for 19 examples. AutoTest, due to its random core, achieves the lowest coverage, reaching exhaustive coverage for only 7 examples.
\begin{table}
\vspace{-0.5cm}
\begin{adjustbox}{width=1\textwidth}
    \begin{threeparttable}
    \caption{\textbf{Overall result}}
\centering
 \renewcommand\arraystretch{1.3}
    \begin{tabular}{ c c c c }
    \hline
      Metrics & SC & IntelliTest & AutoTest
    \\ \hline
    \rowcolor{blue!5} Avg. branch coverage & 99.37\% & 97.15\% & 81.2\%
   \\
   Number of examples reaching exhaustive coverage  & 20 & 19 & 7
   \\
  \rowcolor{blue!5} Avg. time for reaching exhaustive coverage (s) & 0.487 & 27 & 259
  \\
  Avg. number of generated tests for reaching exhaustive coverage & 6.26 & 10.47 & 623.28
   \\ \hline
    \end{tabular}%
    \label{table: overview}%
       \end{threeparttable}
\end{adjustbox}
\vspace{-0.6cm}
\end{table}

To reach exhaustive coverage, SC performs significantly faster than the other two: it needs less than 0.5 seconds on average --- about 50 times less than IntelliTest and 500 times than AutoTest. SC also generates the smallest test suite; the average size of the exhaustive-coverage test suite from IntelliTest is slightly larger than SC, and both are much smaller than AutoTest. The importance of minimizing the size of test suites  has become a crucial concern \cite{orsomints}.

\subsection{Detailed results}

Table \ref{table: coverage} shows coverage results. For each example, we executed the generated test suite and calculated coverage as the ratio of \emph{number of exercised branches} over \emph{number of branches}.
SC always reaches exhaustive coverage (the maximum possible for \e{Lamp} is 87.5\% as it contains an unreachable branch).
IntelliTest reaches exhaustive coverage for most examples but misses it for \e{Account} and \e{Lamp}.
AutoTest's coverage varies from 50\% to 100\%. Occasionally, it performs better than IntelliTest, reaching the maximum 87.5\% for \e{Lamp} against IntelliTest's 50\%.

%it is difficult for AutoTest to satisfy a complex precondition guarding a routine with the random strategy.

%\begin{equation}
%   \e{branch\ coverage} =  \frac{{\e{number\ of\ %exercised\ branches}}}{{\e{number\ of\ branches}}}
%\end{equation}

\begin{table}
\vspace{-1.2cm}
\begin{adjustbox}{width=1\textwidth}
    \begin{threeparttable}
    \caption{\textbf{Result: branch coverage}}
    \begin{subtable}
     \scriptsize
  \centering
   \renewcommand\arraystretch{1.3}
    \begin{tabular}{ c c c c c c p{35pt} p{35pt} p{35pt} p{35pt} c}
    \hline
      & Account  & Clock & Heater  & Lamp & Max & Linear Search & Insertion Sort & Gnome Sort & Square root& Sum and max
     \\ \hline
     \rowcolor{blue!5} SC & 100\% & 100\% & 100\% & 87.5\% & 100\% & 100\% & 100\% & 100\% & 100\% & 100\%
      \\
   IntelliTest & 92.85\% & 100 \% & 100\% & 50\% & 100\% & 100\% & 100\% & 100\% & 100\% & 100\%
   \\
   \rowcolor{blue!5}  AutoTest & 78.6\% & 70\% & 62.5\% & 87.5\% & 66.7\% & 100\% & 80\% & 60\% & 100\% &  100\%
   \\ \hline
    \end{tabular}%
\end{subtable}
\vspace{-0.7cm}

\begin{subtable}
\centering
    \begin{tabular}{ c p{30pt} p{60pt} p{28pt} p{35pt} p{35pt} p{30pt} p{35pt} p{30pt} p{45pt}}
    \hline
     Arithmetic & Binary search & Recursive binary search & Dutch flag & Two way max & Two way sort & Quick sort & Selection Sort & Bubble Sort & Optimized gnome sort
    \\ \hline
    \rowcolor{blue!5} 100\% & 100\% & 100\%& 100\%& 100\%& 100\%& 100\%& 100\%& 100\%& 100\%
   \\
      100\%   & 100\% & 100\% & 100\% & 100\% & 100\% & 100\% & 100\% & 100\% & 100\%
   \\
   \rowcolor{blue!5} 100\%  & 100\% & 85.7 \% & 72.7 \% & 75\% & 83.3\% & 100\% & 80\% & 80\% & 50\%
   \\ \hline
    \end{tabular}%
   \end{subtable}
   \label{table: coverage}%
       \end{threeparttable}
\end{adjustbox}
\vspace{-0.4cm}
\end{table}

\noindent Table \ref{table: duration} gives the time needed to produce the test suite in the various approaches, using the following conventions:
\begin{itemize}[$\bullet$]
  \item For SC, time for test generation includes two parts: proof time (for AutoProof) and time  for extracting tests from failed proofs (time for FP-FT).
  \item For AutoTest, the time is  always the 10-minute timeout, chosen from experience: within that time, test generation of examples usually reaches a plateau.
  \item  IntelliTest does not directly provide time information. We measure duration manually by recording the the timestamps of session start and termination.
\end{itemize}
\noindent In Table \ref{table: duration} results, SC is the fastest of the three, with all its test generation runs taking less than 1 second. For IntelliTest, test generation takes less than 40 seconds for most examples, but three of them out of 20 require more than one minute. For AutoTest, test generation time varies from  1.71 seconds for \e{Square root} to more than 20 minutes for \e{Sum and max}.

\begin{table}
\vspace{-1cm}
\begin{adjustbox}{width=1\textwidth}
    \begin{threeparttable}
    \caption{\textbf{Result: time (in seconds) to reach maximum coverage}}
    \begin{subtable}
     \scriptsize
  \centering
   \renewcommand\arraystretch{1.3}
    \begin{tabular}{ c c c c c c p{35pt} p{35pt} p{35pt} p{35pt} c}
    \hline
      & Account  & Clock & Heater  & Lamp & Max & Linear Search & Insertion Sort & Gnome Sort & Square root& Sum and max
     \\ \hline
    \rowcolor{blue!5} SC & 0.56 & 0.44 & 0.85 & 0.39 & 0.37 & 0.36 & 0.42 & 0.52 & 0.26 & 0.37
   \\
   IntelliTest & 9.58 & 7.44 & 8.06 & -- & 8.19 & 9.63 & 11.77 & 10.89 & 12.86 & 10.99
   \\
   \rowcolor{blue!5} AutoTest & -- & -- & -- & 233.03 & -- & 21.95 & -- & -- & 1.71 &  1322.61
   \\ \hline
    \end{tabular}%
\end{subtable}
\vspace{-0.7cm}

\begin{subtable}
\centering
    \begin{tabular}{c p{30pt} p{60pt} p{25pt} p{35pt} p{35pt} p{30pt} p{35pt} p{30pt} p{45pt}}
    \hline
     Arithmetic & Binary search & Recursive binary search & Dutch flag & Two way max & Two way sort & Quick sort & Selection Sort & Bubble Sort & Optimized gnome sort
    \\ \hline
   \rowcolor{blue!5} 0.415 & 0.44 & 0.48 & 0.43 & 0.52 & 0.39 & 0.90 & 0.50 & 0.59 & 0.54
   \\
    32.98 & 99.29 & 13.07 & 31.36 & 9.59 & 80.91 & 111.57 & 17.81 & 14.74 & 12.32
   \\
    \rowcolor{blue!5} 14.49 & 150.86 & -- & 330.89 & -- & -- & 78.37 & -- & -- & --
   \\ \hline
    \end{tabular}%
   \end{subtable}
   \label{table: duration}%
       \end{threeparttable}
\end{adjustbox}
\vspace{-0.6cm}
\end{table}

\noindent Another important criterion, when a tool covers all the branches of a program, is how many redundant tests it produces. Table \ref{table: size} presents the sizes of the generated test suites of the three tools when reaching exhaustive coverage. From a software engineering viewpoint, particularly for the long-term health of a project, a smaller size achieving the same coverage is better, since it results in a more manageable test suite giving the project the same benefits as a larger one.

Among the three tools, SC generates the fewest tests. In most cases, the number of tests is the same as the number of blocks: as each generated test results from a proof failure of an incorrect instruction, seeded at one program location, each test covers just the corresponding block and introduces no redundancy. If nested branches are present, the size of the test suite can actually be less than the number of branches: SC only generates tests targeting the innermost branches (the leaf nodes of the control structure), as explained in section \ref{nested}; each test going through these branches automatically covers all its enclosing branches.
Intellitest also generates small test suites, but is slower. The reason is Intellitest's use of concolic testing \cite{concolic}, which tests all feasible execution paths: since a a branch can occur in several paths, a test will often identify a branch that was already covered by a different path.
AutoTest, for its part, produces much larger test suites: as an Adaptive Random Testing tool, it often generates multiple test cases covering the same branches.

\begin{table}
\vspace{-1.5cm}
\begin{adjustbox}{width=1\textwidth}
    \begin{threeparttable}
    \caption{\textbf{Result: number of generated tests to reach exhaustive coverage}}
    \begin{subtable}
     \scriptsize
  \centering
   \renewcommand\arraystretch{1.3}
    \begin{tabular}{c c c c c c p{35pt} p{35pt} p{35pt} p{35pt} c}
    \hline
      & Account  & Clock & Heater  & Lamp & Max & Linear Search & Insertion Sort & Gnome Sort & Square root& Sum and max
     \\ \hline
     \rowcolor{blue!5}SC & 13 & 10 & 8 & 7 & 3 & 3 & 3 & 3 & 4 & 3
   \\
   IntelliTest & 13 & 13 & 8 & -- & 4 & 7 & 5 & 7 & 5& 5
   \\
   \rowcolor{blue!5}AutoTest & -- & -- & -- & 656 & -- & 127 & -- & -- & 18 & 1784
   \\ \hline
    \end{tabular}%
\end{subtable}
\vspace{-0.7cm}
\begin{subtable}
\centering
    \begin{tabular}{c p{30pt} p{60pt} p{25pt} p{35pt} p{35pt} p{30pt} p{35pt} p{30pt} p{45pt}}
    \hline
     Arithmetic & Binary search & Recursive binary search & Dutch flag & Two way max & Two way sort & Quick sort & Selection Sort & Bubble Sort & Optimized gnome sort
    \\
   \rowcolor{blue!5}  14 & 4 & 7 & 9 & 2 & 5 & 9 & 5 & 4 &  7
   \\
    25 & 6 & 15 & 27 & 4 & 9 & 18 & 12 & 8 & 8
   \\
    \rowcolor{blue!5} 531 & 905 & -- & -- & -- & -- & 342 & -- & -- & --
   \\ \hline
    \end{tabular}%
   \end{subtable}
   \label{table: size}%
       \end{threeparttable}
\end{adjustbox}
\vspace{-0.6cm}
\end{table}

\noindent Tables \ref{table: overview} to \ref{table: size} provide evidence of the benefits of the approach (subject to the limitations examined in the next section):
SC is fast and efficient; it uses less than 1 second to produce an exhaustive-coverage test suite with the fewest number of test cases. Other observations:
\begin{itemize} [$\bullet$]
    \item AutoTest does not guarantee that the test inputs satisfy the routine's precondition, while SC and IntelliTest always generate precondition-satisfying test inputs. The reason is that SC and IntelliTest rely on  the results of constraint solving, where the routine's precondition is encoded as an assumption and will always be satisfied.
    \item The SC approach is has a prerequisite: the program under test has to be proved correct (the proof of the original program has no failure), while AutoTest and IntelliTest have no such constraint.
    \item As to the values of the generated test inputs, IntelliTest and AutoTest  always apply small values that are easy to understand. SC initially produces test inputs that may contain large values; its ``minimization'' mechanism \cite{huang2022improving} corrects the problem.
\end{itemize}

\section{Limitations and threats to validity} \label{limitations}
The setup of the SC approach assumes a Hoare-style verification framework (of which Boogie is but one example), and the availability of a test generation mechanism that supports generating test cases from proof failures. We have not studied the possible application of the ideas to different verification frameworks, based for example on abstract interpretation or model checking.

The current version of SC is subject to the following limitations:
\begin{itemize}[$\bullet$]
    \item SC is not able to handle programs with non-linear computations (such as derivation and exponentiation); this restriction comes from the underlying SMT solver. 
    \item SC does not support the more advanced parts of the Eiffel system, in particular generic classes. Data structures are limited to arrays and sequences. 
\end{itemize}
These limitations will need to be removed for SC to be applicable to industrial-grade programs.

The following considerations may influence the generalization of the results achieved so far:
\begin{itemize}[$\bullet$]
    \item The number of repeated experiments increased the potential threats to internal validity. We hope that further experiments with large number of iterations will provide more conclusive evidence.
    \item Although a few of the examples classes that we processed so far are complex and sophisticated, most are of a small size and not necessarily representative of industrial-grade object-oriented programs. In the future, we intend to use the EiffelBase library\footnote{EiffelBase Data Structures: \url{https://www.eiffel.org/doc/solutions/EiffelBase_Data_Structures_Overview}}, which has yielded extensive, representative results in the evaluation of AutoProof and AutoTest, and exhibits considerable variety and complexity in terms of size (according to various metrics), richness of program semantics, and sophistication of algorithms and software architecture.
\end{itemize}

\section{Related work} \label{related}

%To our knowledge, the core idea of this article --- to produce a test guaranteed to cover any given branch of a program, seed an incorrect instruction into it, run a proof that fails, and collect the corresponding automatically-generated counterexample --- is new.

Previous work has taken advantage of counterexamples generated by failing proofs, but for other purposes, in particular automatic program repair \cite{nilizadeh2021more} and generation of failing tests \cite{nilizadeh2022generating} \cite{huang2022failed}. These techniques work on the original program and not, as here, on a transformed program in which \textit{incorrect} instructions have been inserted with the express purpose of making the proof fail.

%\li{To idea of this article--- to produce a test guaranteed to cover any given branch of a program, seeding incorrect instruction into it, run a proof that fails, and collect the corresponding automatically-generated counterexample --- is not completely new. }

The earliest work we know to have applied this idea \cite{angeletti2009improving} \cite{angeletti2009automatic} generates tests for low-level C programs using Bounded Model Checking (BMC) \cite{kroening2014cbmc}, producing test suites with exhaustive branch coverage. A more recent variant, for Java bytecode, is JBMC \cite{brenguier2023jbmc}. In contrast with SC, each verification run  only activates one assertion at a time, producing one counterexample. This approach is conceptually similar, in the terminology of the present work (\ref{variables}), to the ``MSP'' (Multiple Seeded Programs) technique, although the C version \cite{angeletti2009improving} uses compile-time macros, one for each block, to avoid the actual generation of multiple programs. In contrast, the present work uses RSSP (Repeatedly Seeded Single Program), relying on a single \textit{run-time} variable representing the block number.  BMC-based approaches rely on the correctness of the \textit{bound} of the execution trace: if the bound is not set correctly, some branches might not be covered, requiring more verification runs to obtain a better bound.

Other techniques that apply constraint solving for generating inputs includes test generations based on symbolic execution, such as Pex/IntelliTest \cite{tillmann2008pex}, KLEE \cite{cadar2008klee}, PathCrawler\cite{williams2021towards}. None of the strategies proposed guarantees exhaustive branch coverage; they can  achieve it when a systematic test generation strategy, rather than one based on heuristics or randomization, is applied.

A very recent development (published just as the present work was being submitted) is DTest, a toolkit \cite{fedchin2023toolkit} for generating unit tests for Dafny programs, applying ideas similar to those of SC. As the generated Dafny tests are not directly executable, test generation requires transformation of Dafny programs and tests into a mainstream language. In contrast, the present approach works directly on Eiffel programs. The DTest coverage results cited in the referenced article are  100\% on only 2 of its examples, and go down to as low as 58\% on the others. One should not draw definite conclusions from these figures, since the examples are different, their program sizes too (more precisely, most of the examples are of comparable sizes, but the cited work has three between 1100 and 1900 LOCs, which we have not handled yet), and the article does not mention any presence of unreachable code (which makes it impossible to distinguish between full coverage and exhaustive coverage). It should be noted, however, that the article also makes no mention of the ``Seeded Unreachability'' issue discussed in section \ref{variables}; in fact, it states that   ``\textit{DTest enters a loop where it systematically injects trivially failing trap assertions
(meaning assert false)}'', a technique which generally leads, for any program with a non-trivial control structure, to Seeded Unreachability and hence to decreased coverage. That omission may be the reason for the relatively low coverage results reported in the article. The Conditional Seeding technique of SC, introduced by the present work, addresses Seeded Unreachability and has made it possible to reach exhaustive coverage in all examples so far. In addition, to obtain small test suites, DTest seems to require a separate minimization strategy, which takes from 8 to 1860 seconds on the cited examples, far beyond the times of running SC. In discussing minimization, the authors appear to come close to recognizing the Seeded Unreachability issue, without using the Conditional Seeding technique, when they write that ``\textit{we determine the feasibility
of a path via a query to the SMT solver, in which a trap assertion is added
that fails only if all the blocks along the path are visited}'', a technique that is ``\textit{exponential in the number of SMT queries (running on all benchmarks} [cited in the article] \textit{would take weeks)}''. SC does not appear to need any such technique.

%do they guarantee full branch coverage?

%\section{Threat to validity}

\section{Conclusions and future work} \label{conclusion}
The approach presented here, Seeding Contradiction (SC), automatically generates test suites that achieve exhaustive branch coverage very fast. The presentation of the approach comes with a proof of correctness, defined as the guarantee that the generated test suite achieves exhaustive coverage (full coverage of reachable branches). While technical limitations remain, the evaluation so far demonstrates the effectiveness and efficiency of the SC approach through the comparison with two existing test generators IntelliTest and AutoTest, in terms of achieved coverage, generation time, and size of the test suite.

Ongoing work includes handling larger examples, processing entire classes instead of single routines, providing a mechanism to generate tests covering branches that a given test suite fails to cover, and taking advantage of the SC strategy to identify dead code.

\vspace{0.5cm}
\noindent \textbf{Acknowledgements}. We are particularly grateful, for their extensive and patient help, to  Yi Wei (AutoTest) and Jocelyn Fiat (EiffelStudio and AutoProof). The paper benefitted from perceptive comments by the anonymous referees on the original version. 

\bibliographystyle{splncs04}
\bibliography{reference}
%(12 to 15 pages, plus at most 2 extra pages for references in the one-column Springer LNCS format).

\end{document}